\documentclass[12pt,superscriptaddress,aps,prd,preprint]{revtex4}
\usepackage{amsmath}
\usepackage{amssymb}
\makeatletter
\newcommand{\bea}{\begin{eqnarray}}
\newcommand{\eea}{\end{eqnarray}}

\newcommand{\pa}{\partial}

\usepackage[dvips]{graphicx}

\begin{document}
\title{Dynamical Chern-Simons modified gravity and Friedmann-Robertson-Walker metric}
\author{C. Furtado}
\affiliation{Departamento de F\'{\i}sica, Universidade Federal da Para\'{\i}ba,\\
Caixa Postal 5008, 58051-970, Jo\~ao Pessoa, Para\'{\i}ba, Brazil}
\email{furtado,jroberto,petrov@fisica.ufpb.br}
\author{J. R. Nascimento}
\affiliation{Departamento de F\'{\i}sica, Universidade Federal da Para\'{\i}ba,\\
Caixa Postal 5008, 58051-970, Jo\~ao Pessoa, Para\'{\i}ba, Brazil}
\email{furtado,jroberto,petrov@fisica.ufpb.br}
\author{A. Yu. Petrov}
\affiliation{Departamento de F\'{\i}sica, Universidade Federal da Para\'{\i}ba,\\
Caixa Postal 5008, 58051-970, Jo\~ao Pessoa, Para\'{\i}ba, Brazil}
\email{furtado,jroberto,petrov@fisica.ufpb.br}
\author{A. F. Santos}
\affiliation{Instituto de F\'{\i}sica, Universidade Federal de Mato Grosso\\
78060-900, Cuiab\'{a}, Mato Grosso, Brazil}
\email{alesandroferreira@fisica.ufmt.br}

\begin{abstract}
We study the conditions for the consistency of the Friedmann-Robertson-Walker (FRW) metric with the dynamical Chern-Simons modified gravity. It turns out to be that in this situation the accelerated expansion of the Universe takes place, with the time dependence of the scale factor turns out to be similar to the case of presence of the Chaplygin gas. Also we found that this modification changes the total density of the Universe and therefore gives a nontrivial impact to a cosmological scenario.
\end{abstract}

\maketitle

Studies of the gravity, without doubts, can be treated as one of the most important lines in the modern theoretical physics. The interest to the gravity highly uprose because of the discovery of the accelerated expansion of the Universe \cite{Riess}, which, together with another key problem of the gravity theory, that is, the search for the consistent quantum description of the gravity, called the attention to construction of the modified gravity models. The idea of possibility for the breaking the Lorentz and CPT symmetries implied in a high interest to the Lorentz-CPT breaking modifications of the gravity, whose wide list is given in \cite{Kost}. The most interesting and popular of these extensions is certainly the four-dimensional gravitational Chern-Simons term originally introduced in \cite{JaPi}. Many aspects of this term are discussed in \cite{Yunes} (see also \cite{ours} for a perturbative generation of this term).
 
The further development of the studies of the four-dimensional gravitational Chern-Simons term implied in a formulation of the Chern-Simons modified gravity whose action is a sum of the usual Einstein-Hilbert action and of the gravitational Chern-Simons term, and, further, in a concept of the dynamical Chern-Simons coefficient initially motivated by the fact that, while many metrics, including the spherically and cylindrically symmetric ones \cite{Grumiller} and the G\"{o}del one \cite{Go} solve the modified Einstein equations, the Kerr metric does not. It was shown in \cite{Grumiller,Erick,Konno} that the Kerr metric, known as one of the most important solutions of the usual Einstein equations, can solve the modified ones only if, first, this metric is modified itself, second, the Chern-Simons coefficient acquires a nontrivial dynamics. As a result, the natural problem arises -- whether other known metrics solve the equations of motion in the case of the dynamical Chern-Simons coefficient? In this paper, we consider this problem in the case of the Friedmann-Robertson-Walker (FRW) metric. Furthermore, we estimate the impact of the dynamical Chern-Simons coefficient to the expansion of the Universe.
 
The starting point of our consideration is the following action for the Chern-Simons modified gravity:
\bea
\label{smod}
S=\frac{1}{16\pi G}\int d^4x\Bigl[\sqrt{-g}R
+\frac{l}{4}\Theta \,{^*}RR-\frac{1}{2}\pa^\mu\Theta\pa_\mu\Theta\Bigl] + S_{mat},
\eea
where, unlike \cite{Go}, the function $\Theta$ is a dynamical variable \cite{Erick}.  
Varying this action with respect to the metric and to the scalar field $\Theta$, we obtain the following equations of motion:
\bea
&& G_{\mu\nu}+lC_{\mu\nu}= 8\pi G T_{\mu\nu}, \\
&&g^{\mu\nu}\nabla_{\mu}\nabla_{\nu}\Theta=-\frac{l}{64\pi} {^*}RR,
\label{einst}
\eea
where $G_{\mu\nu}$ is the Einstein tensor (within this formulation, the contribution from the cosmological term is absorbed into the energy-momentum tensor) and $C_{\mu\nu}$ is the Cotton tensor defined as
\bea
C^{\mu\nu}=-\frac{1}{2\sqrt{-g}}\Bigl[v_\sigma \epsilon^{\sigma\mu\alpha\beta}D_\alpha R^\nu_\beta+\frac{1}{2}v_{\sigma\tau}\epsilon^{\sigma\nu\alpha\beta}R^{\tau\mu}\,_{\alpha\beta}\Bigl] \,+\, (\mu\longleftrightarrow\nu), \label{cotton}
\eea
where $v_{\sigma}=D_{\sigma}\Theta$, $v_{\sigma\tau}=D_{\sigma}D_{\tau}\Theta$, and ${^*}RR$ is the topological invariant called the Pontryagin term whose explicit form is
\bea
{^*}RR\equiv {^*}{R^a}\,_b\,^{cd}R^b\,_{acd},
\eea
where $R^b\,_{acd}$ is the Riemann tensor and ${^*}{R^a}\,_b\,^{cd}$ is the dual Riemann tensor given by
\bea
{^*}{R^a}\,_b\,^{cd}=\frac{1}{2}\epsilon^{cdef}R^a\,_{bef}.
\eea
In this paper we are concentrated on the case of the Friedmann-Robertson-Walker metric which is written as
\bea
ds^2=- dt^2+a(t)^2\left[\frac{dr^2}{1-kr^2}+r^2d\theta ^2+r^2\sin ^2\theta d\phi ^2\right],
\eea
where $a(t)$ is the scale factor.

Using the Christoffel symbols for this metric
\begin{eqnarray}
\Gamma^0_{11}&=&\frac{a\dot{a}}{1-kr^2}, \quad\,
\Gamma^0_{22}=a \dot{a}r^2, \quad\,
\Gamma^1_{01}=\Gamma^2_{02}=\Gamma^3_{03}=\frac{\dot{a}}{a},\nonumber\\
\Gamma^1_{22}&=&-r(1-kr^2),\quad\,
\Gamma^2_{12}=\Gamma^3_{13}=\frac{1}{r},\quad\,
\Gamma^2_{33}=-\sin\theta\cos\theta,\nonumber\\
\Gamma^1_{11}&=&\frac{kr}{1-kr^2}, \quad\,
\Gamma^0_{33}=a\dot{a}r^2\sin^ 2 \theta, \quad\,
\Gamma^1_{33}=-r(1-kr^2)\sin^ 2\theta, \quad\,
\Gamma^3_{23}=\rm{cot\theta},
\end{eqnarray}
one can find that the nonzero components of the Riemann tensor are
\begin{eqnarray}
R_{0101}&=&-\frac{a\ddot{a}}{1-kr^2},\quad\,
R_{0202}=-a\ddot{a}r^2,\quad\,
R_{0303}=-r^2a\ddot{a}\sin^2\theta,\\
R_{1212}&=&\frac{a^2r^2(\dot{a}^2+k)}{1-kr^2},\quad\,
R_{1313}=\frac{r^2a^2\sin^ 2\theta(\dot{a}^2+k)}{1-kr^2},\quad\,
R_{2323}=a^2r^4\sin^ 2\theta(\dot{a}^ 2+k),\nonumber
\end{eqnarray}
the nonzero components of the Ricci tensor are
\begin{eqnarray}
R_{00}&=&-3\frac{\ddot {a}}{a},\quad\,
R_{11}=\frac{a\ddot{a}+2\dot{a}^2+2k}{1-kr^2},\nonumber\\
R_{22}&=&r^2(a\ddot{a}+2\dot{a}^2+2k),\quad\,
R_{33}=r^2(a\ddot{a}+2\dot{a}^2+2k)\sin ^2\theta,
\end{eqnarray}
and the Ricci scalar is 
\begin{eqnarray}
R=6\left[\frac{\ddot{a}}{a}+\left(\frac{\dot{a}}{a}\right)^2+\frac{k}{a^2}\right].
\end{eqnarray}
Let us now consider the modified Einstein equations (\ref{einst}). First of all, we can study the Cotton tensor (\ref{cotton}). Calculating component by component we obtain that
\bea
C^{\mu\nu}=0,
\eea
i.e.,  the Cotton tensor completely vanishes for the FRW metric, at any $\Theta(t,r,\theta,\phi)$.

It is easy to see that the Pontryagin invariant also vanishes for the FRW metric. Indeed,
\bea
{^*}RR\equiv \frac{1}{2}\epsilon^{cdef}R^a\,_{bef}R^b\,_{acd},
\eea
but, as we already noted, the only non-zero components of the Riemann tensor for the FRW metric have the form $R_{[ab][ab]}$. Hence sets of indices $ab$, $cd$, $ef$ must coincide, and this condition annihilates the Levi-Civita tensor immediately. All this confirms the statement that the Chern-Simons term does not contribute to the cosmological part of the equations of motion \cite{Soda}. However, the impact of the fact that the Chern-Simons coefficient possesses a nontrivial dynamics is essential for the energy-momentum tensor in the theory. Also, we note that in our case, unlike \cite{Soda}, we have both the scalar field $\Theta$ and the usual matter whose contributions to the energy-momentum tensor will be given further.

Therefore, the equations of motion are reduced to
\bea
&& G_{\mu\nu}= 8\pi GT_{\mu\nu}, \\
&&g^{\mu\nu}\nabla_{\mu}\nabla_{\nu}\Theta=0.
 \label{einst1}
\eea

We choose the energy-momentum tensor $T_{\mu\nu}$ to be composed of two terms: 
\bea
T_{\mu\nu}=T_{\mu\nu}^M+T_{\mu\nu}^{\Theta}, 
\eea
where $T_{\mu\nu}^M$ is the energy-momentum tensor of an usual matter, and the energy-momentum tensor for the $\Theta$ field looks like (cf. \cite{Konno})
\bea
T_{\mu\nu}^{\Theta}=(\partial_{\mu}\Theta)(\partial_{\nu}\Theta)-\frac{1}{2}g_{\mu\nu}(\partial^{\lambda}\Theta)(\partial_{\lambda}\Theta),
\eea
that is, it reproduces the form of the usual energy-momentum tensor for the scalar field in a curved space. Solving of these equations of motion seems to be no more difficult as in the case of absence of $\Theta$. 

To proceed, we can find that the components of the Einstein tensor $G_{\mu\nu}=R_{\mu\nu}-\frac{1}{2}Rg_{\mu\nu}$ look like
\bea
G_{00}=\frac{3}{a^2}(\dot{a}^2+k), \quad \, G_{ij}=-\frac{1}{a^2}(2a\ddot{a}+\dot{a}^2+k)g_{ij}.
\eea
Let us suppose for the sake of the simplicity (this choice also satisfies the principles of the space homogeneity and isotropy) that $\Theta=\Theta(t)$. We can rewrite the equation (\ref{einst1}) as
\bea
g^{\mu\nu}\nabla_{\mu}\nabla_{\nu}\Theta=g^{\mu\nu}(\partial_{\mu}\partial_{\nu}-\Gamma_{\mu\nu}^{\lambda}\partial_{\lambda})\Theta=0.
\eea
Taking into account that $g^{ij}\Gamma^0_{ij}=3\frac{\dot{a}}{a}$, and that $\Theta$ depends only on $t$, we find that
\bea
\label{theta}
\ddot{\Theta}+3\frac{\dot{a}}{a}\dot{\Theta}=0.
\eea 
Hence, $\Theta$ is related to $a$ as
\bea
\dot{\Theta}=Ca^{-3}.
\eea

The energy-momentum tensor for the $\Theta$ field hence is given by
\bea
T^{\Theta}_{00}&=&\frac{1}{2}\dot{\Theta}^2=\frac{C^2}{2}a^{-6},\nonumber\\
T^{\Theta}_{ij}&=&\frac{1}{2}g_{ij}\dot{\Theta}^2=\frac{C^2}{2}g_{ij}a^{-6}.
\eea
The modified Einstein equations, in the relevant case $\mu=\nu=0$, looks like
\bea
\label{frie}
\frac{3}{a^2}\left(\dot{a}^2+k\right)=\frac{8\pi G}{3}\left(\frac{C^2}{2}a^{-6}+T^M_{00}\right).
\eea
Here $T^M_{\mu\nu}$ is an energy-momentum tensor of the usual matter (apart from the $\Theta$ field).

Also, we must take into account the equation of conservation of energy, that is given to the component $\nu=0$ of the equation (cf. \cite{carroll})
\bea
\label{mem}
\nabla_{\mu}T^{\mu\nu}=0.
\eea

Here $T^{\mu\nu}$ is composed both by the contributions from the $\Theta$ field and from the usual matter which by our choice is described by a perfect fluid. Thus the energy-momentum tensor of the matter is
\bea
T_{\mu\nu}^M=(\rho+p)U_\mu U_\nu +pg_{\mu\nu}, 
\eea
where $U_\mu=(1,0,0,0)$ is the four-velocity of the fluid. Substituting this expression to (\ref{mem}), we find that the equation of conservation of energy (\ref{mem}) in the modified Einstein gravity with a dynamical Chern-Simons coefficient, in the relevant $00$ sector, is given by
\bea
\partial_0T^{00}+\Gamma^{\mu}_{\mu 0}T^{00}+\Gamma^0_{\mu\lambda}T^{\mu\lambda}=0,
\eea
which, since $\Gamma^{\mu}_{\mu 0}=3\frac{\dot{a}}{a}$, $T_{00}=\rho+\frac{1}{2}\dot{\Theta}^2$, $\Gamma^0_{ij}=\frac{1}{2}\partial_0 g_{ij}=\frac{\dot{a}}{a}g_{ij}$ and $T^{ij}=(p+\frac{1}{2}\dot{\Theta}^2)g^{ij}$, thus, $\Gamma^0_{ij}T^{ij}=\Gamma^0_{ij}g^{ij}(p+\frac{1}{2}\dot{\Theta}^2)=3\frac{\dot{a}}{a}(p+\frac{1}{2}\dot{\Theta}^2)$, yields
\bea
\label{leq}
\frac{d}{dt}\left(\rho+\frac{C^2}{2}a^{-6}\right)+3\frac{\dot{a}}{a}\left(\rho+p+C^2a^{-6}\right)=0. 
\eea
We note that in the case $\Theta=const$ (that is, $C=0$), where the gravitational Chern-Simons term is reduced to an irrelevant total derivative, we recover the usual result (see f.e. \cite{carroll}).
The only impact of the dynamical Chern-Simons term in this equation arices in the presence of $C^2$.
If we suppose that the matter in the Universe is composed by one dominant component and obeys the linear equation of state,
\bea
p=(\gamma-1)\rho,
\eea
where $\gamma$ is a numerical factor characterizing the nature of the matter (it is equal to 1 for the dust, $4/3$ for the radiation and 0 for the pure cosmological term), we can rewrite (\ref{leq}) as
\bea
\label{leq1}
\frac{d}{dt}\left(\rho+\frac{C^2}{2}a^{-6}\right)+3\frac{\dot{a}}{a}\left(\gamma\rho+C^2a^{-6}\right)=0. 
\eea
It remains to solve the following system (we choose the case $k=0$):
\bea
&&\frac{\dot{a}^2}{a^2}=\frac{8\pi G}{3}\left(\rho+\frac{C^2}{2}a^{-6}\right);\nonumber\\
&&\frac{d}{dt}\left(\rho+\frac{C^2}{2}a^{-6}\right)+3\frac{\dot{a}}{a}\left(\gamma\rho+C^2a^{-6}\right)=0.
\eea
This system can be simplified. Indeed, it is easy to see that $\frac{d}{dt}(\frac{C^2}{2}a^{-6})+3\frac{\dot{a}}{a}C^2a^{-6}=0$, hence we find
\bea
&&\frac{\dot{a}^2}{a^2}=\frac{8\pi G}{3}\left(\rho+\frac{C^2}{2}a^{-6}\right);\nonumber\\
&&\frac{d\rho}{dt}+3\frac{\dot{a}}{a}\gamma\rho=0.
\eea
It follows from the last equation of this system that the density and the scale factor are related as
\bea
\rho=\rho_0a^{-3\gamma},
\eea
thus, we arrive at the following Friedmann equation (modified)
\bea
\frac{\dot{a}^2}{a^2}=\frac{8\pi G}{3}\left(\rho_0a^{-3\gamma}+\frac{C^2}{2}a^{-6}\right).
\eea
The formal solution of this equation is
\bea
t=\int\frac{da a^z}{\sqrt{A+Ba^{2z-4}}},
\eea
where $z=\frac{3\gamma}{2}-1$, $A=\frac{8\pi G\rho_0}{3}$, $B=\frac{4\pi G C^2}{3}$.
Evaluating this expression, we arrive at
\bea
{}_2F_1\left(\frac{1}{2-\gamma},\frac{1}{2};\frac{\gamma-3}{\gamma-2};-2\frac{\rho_0}{C^2}a^{6-3\gamma}\right)a^3(t)=3C\sqrt{\frac{4\pi G}{3}}t.
\eea 
The similar solution has been obtained for the Chaplygin gas which is known to be one of the possible mechanisms providing an accelerated expansion of the Universe \cite{costa}.

Now, let us consider the situation where the cosmological constant differs from zero, and matter, beside of the cosmological constant and $\Theta$ field, is composed by two or more distinct components. The most important example is the Universe involving both dust-like matter and radiation.
In this case, the modified Friedmann equation looks like
\bea
\frac{\dot{a}^2}{a^2}+\frac{k}{a^2}=\frac{8\pi G}{3}\sum\limits_i \rho_{0i}a^{-3\gamma_i}+\frac{\Lambda}{3}+\frac{4\pi G}{3}C^2a^{-6},\label{53}
\eea
We can introduce the Hubble parameter $H=\frac{\dot a}{a}$ and the density parameters for different types of matter $\Omega=\frac{\rho}{\rho_c}$,
where the critical density $\rho_c$ is $\rho_c=\frac{3H^2}{8\pi G}$. In particular, for the matter we have the relative densities $\Omega_{mi}=\frac{\rho_{mi}}{\rho_c}$, with $\rho_{mi}$ is a density of the matter of type $i$, and for the cosmological constant -- $\Omega_{\Lambda}=\frac{\rho_{\Lambda}}{\rho_c}$, with $\rho_{\Lambda}=\frac{\Lambda}{8\pi G}$. Then, we define a density of the $\Theta$ field as $\rho_{\Theta}=\frac{C^2a^{-6}}{2}$,
with the density parameter is $\Omega_{\Theta}=\frac{\rho_{\Theta}}{\rho_c}=\frac{4\pi G}{3H^2}C^2a^{-6}$. Finally, the equation (\ref{53}) takes the form
\bea
1+\frac{k}{a^2H^2}=(\sum\limits_i\Omega_{mi})+\Omega_{\Lambda}+\Omega_{\Theta}.
\eea
Also we define the density parameter associated to the curvature as
\bea
\frac{k}{a^2H^2}\equiv -\Omega_k,
\eea
thus, the sum of all relative densities is 1 as it must be:
\bea
(\sum\limits_i\Omega_{mi})+\Omega_{\Lambda}+\Omega_k+\Omega_{\Theta}=1.\label{omegas}
\eea
If $\Omega_{\Theta}=0$ we recuperate the usual result \cite{Liddle}. Thus, Chern-Simons modified gravity with a dynamic $\Theta$ field yields one more contribution to the total density of the Universe. 

Now, let us divide the equation (\ref{53}) by the Hubble constant today $H_0^2$. Suggesting that the Universe, besides of cosmological constant, curvature and the dynamic scalar field involves two types of the usual matter, that is, dust ($\gamma=1$) and radiation ($\gamma=4/3$), we have that
\bea
\left(\frac{H}{H_0}\right)^2+\frac{k}{a^2H_0^2}=\frac{8\pi G\rho_{0,m}}{3H_0^2}a^{-3}+\frac{8\pi G\rho_{0,r}}{3H_0^2}a^{-4}+\frac{\Lambda}{3H_0^2}+\frac{4\pi G}{3H_0^2}C^2a^{-6},
\eea
where $\rho_{0,d}$ and $\rho_{0,r}$ are the actual densities of dust and radiation respectively. In terms of the density parameters 
\bea
\left(\frac{H}{H_0}\right)^2+\frac{k}{a^2H_0^2}=\frac{\Omega_{0,d}}{a^3}+\frac{\Omega_{0,r}}{a^4}+\Omega_{0,\Lambda}+\Omega_{0,\Theta}.\label{67}
\eea

Coming back to equation (\ref{omegas}) we can obtain
\bea
\frac{k}{a^2H^2}&=&\Omega_d+\Omega_r+\Omega_\Lambda+\Omega_\Theta - 1=\Omega_0-1,
\eea
where $\Omega_0=\Omega_d+\Omega_r+\Omega_\Lambda+\Omega_\Theta$. To fix the initial conditions, we note that for the actual phase of the Universe we have
\bea
\frac{k}{a_0^2H_0^2}=\Omega_0-1.
\eea
Thus the equation (\ref{67}) becomes
\bea
\frac{H}{H_0}=\left[\frac{\Omega_{0,d}}{a^3}+\frac{\Omega_{0,r}}{a^4}+\Omega_{0,\Lambda}+\Omega_{0,\Theta}+\frac{1-\Omega_0}{a^2}\right]^{\frac{1}{2}},
\eea
with $a_0=1$ being the actual value of the scale factor. Integrating this equation, we can calculate the age of the Universe 
\bea
H_0t=\int_0^a\frac{a'da'}{\left[\Omega_{0,r}+\Omega_{0,d}a'+(\Omega_{0,\Lambda}+\Omega_{0,\Theta})a'^4+(1-\Omega_0)a'^2\right]^{\frac{1}{2}}}.
\eea
We see that the age of the Universe in our model depends of the new contribution $\Omega_{0,\Theta}$.  

Now, let us estimate a numerical value for $\Omega_{0,\Theta}$. To do it, we consider the particular case $\Omega_{0,r}\approx0$ (that is, the usual dust-like matter dominates over the radiation) and $\Omega_0\approx1$ (approximately flat Universe). Thus
\bea
t=\frac{1}{H_0}\int_0^a\frac{a'da'}{\left[\Omega_{0,d}a'+(\Omega_{0,\Lambda}+\Omega_{0,\Theta})a'^4\right]^{\frac{1}{2}}},
\eea
which gives a result
\bea
t&=&\frac{1}{H_0}\frac{2}{3\sqrt{\Omega_{0,\Lambda}+\Omega_{0,\Theta}}}\log\Big[2\Big(a^{3/2}(\Omega_{0,\Lambda}+\Omega_{0,\Theta})+\nonumber\\&+&\sqrt{\Omega_{0,\Lambda}+\Omega_{0,\Theta}}\sqrt{a^3(\Omega_{0,\Lambda}+\Omega_{0,\Theta})+\Omega_{0,d}}\Big)\Big].
\eea
Now we take into account that for the present Universe $a=a_0=1$, $\Omega_{0,\Lambda}\approx0,7$, $\Omega_{0,d}\approx0,3$, $H_0\approx2,27\times 10^{-18}\, s^{-1}$ and $t_0\approx3,95\times10^{17}s\approx12,5\times10^9\,years$. Therefore we have
\bea
\log\left[2\left(\Omega_{0,\Theta}+\sqrt{\Omega_{0,\Theta}+0,7}\sqrt{\Omega_{0,\Theta}+1}+0,7\right)\right]=1,34223\sqrt{\Omega_{0,\Theta}+0,7},\label{trans}
\eea
This equation can be solved only numerically. The solutions are shown in the Figure 1.         
\begin{figure}[h]
\includegraphics[scale=0.9]{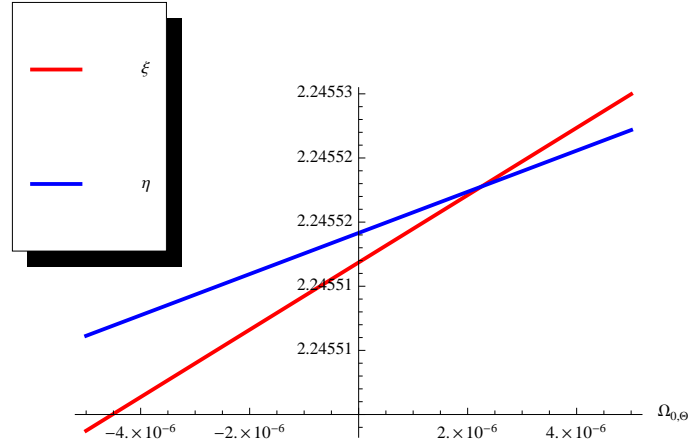}
\caption{Here $\xi=\log\left[2\left(\Omega_{0,\Theta}+\sqrt{\Omega_{0,\Theta}+0,7}\sqrt{\Omega_{0,\Theta}+1}+0,7\right)\right]$ and $\eta=1,34223\sqrt{\Omega_{0,\Theta}+0,7}.$ }
\end{figure}

From this figure we can estimate a numerical value of $\Omega_{0,\Theta}$ as
\bea
\Omega_{0,\Theta}\approx2,2\times10^{-6}.
\eea
Thus we see that in the Universe dominated by a dark matter and a dark energy the presence of the contribution $\Omega_{0,\Theta}$ is small in comparison with other components of the Universe. 

Let us discuss our results. Our key conclusion, with no doubts, is that the modified Einstein gravity with a dynamical Chern-Simons coefficient also admits the FRW solutions, as well as the case where the Chern-Simons coefficient $\Theta$ is treated as an external nondynamical field, see \cite{Grumiller}. Moreover, we show that, at the certain interval of values of $\gamma$ parametrizing the equation of state the scale factor displays an accelerated growth whose scenario is similar to that one emerging within the context of an essentially physically different mechanism, that is, the Chaplygin gas \cite{Cha}.

We also show that in this modified theory there is one more contribution to the matter density, that is, the density parameter of the Universe associated with the dynamical field $\Theta$. Using the known parameters of the Universe, we estimate a numerical value for this parameter which turns to be much more small than the value of density parameter of radiation today, $\Omega_ {0, r} \approx8, 4 \times10 ^{-5}$ \cite{Ryden} (where we find $\Omega_{0,\Theta}\approx2,2\times10^{-6}$), i.e., this contribution, as well as the radiation, does not jeopardize the actual dominance of the dark matter and the dark energy in our Universe.

{\bf Acknowledgments.}
This work was partially supported by Conselho Nacional de
Desenvolvimento Cient\'\i fico e Tecnol\'ogico (CNPq) and Coordena\c c\~ao de Aperfei\c coamento de Pessoal de N\'\i vel Superior (CAPES: AUX-PE-PROCAD 579/2008). A. Yu. P. has been supported by the CNPq project No. 303461-2009/8, and A. F. S. has been supported by the CNPq project, No. 473571/2010-2.

\end{document}